# Influence of Magnetic Anisotropy on the Ground State of [CH$_3$NH$_3$]Fe(HCOO)$_3$: Insights into the Improper Modulated Magnetic Structure

*Laura Cañadillas-Delgado,*[a] *Lidia Mazzuca,*[a] *Sanliang Ling,*[b] *Matthew J. Cliffe*[c] *and Oscar Fabelo*[a]

[a] Institut Laue-Langevin, 71 Avenue des Martyrs, CS 20156, 38042 Grenoble Cedex 9, France.

[b] Advanced Materials Research Group, Faculty of Engineering, University of Nottingham, University Park, Nottingham, NG7 2RD, UK.

[c] School of Chemistry, University Park, University of Nottingham, NG7 2RD, UK.

## Abstract

The hybrid perovskites [CH$_3$NH$_3$]Co$_x$Ni$_{x-1}$(HCOO)$_3$ with x = 0, 0.25, 0.5, 0.75 and 1.0 possess multiple phase transitions including incommensurate structures. Further to this, [CH$_3$NH$_3$]Ni(HCOO)$_3$ has been found to have a proper magnetic incommensurate structure in its ground state. We therefore carry out a detailed structural characterization of the isomorphous [CH$_3$NH$_3$]Fe(HCOO)$_3$ (**1**) to investigate whether it also has incommensurate structural and magnetic modulations. We confirm that **1** crystallizes in the *Pnma* space group at room temperature (RT) with a perovskite structure. Upon cooling, at about 170 K, the occurrence of new satellite reflections in the diffraction pattern show a phase transition to a modulated structure, which could be refined in the *Pnma*(00$\gamma$)0*s*0 super space group with **q**$_1$ = 0.1662(2)$c^*$. On further cooling to 75 K the satellite reflections become closer to the main reflections, indicating a new phase transition that keeps the super space group invariant but changes the modulation wave vector, **q**$_2$ = 0.1425(2)$c^*$, i.e. with a longer modulation period. The structure then does not change structural phase down to base temperature (2 K). Magnetic susceptibility measurements collected under field-cooled and zero-field-cooled reveal a 3D antiferromagnetic order below 17 K. The overlapping in temperature between structural modulation and long-range magnetic order presents a unique opportunity to study magneto-structural coupling. Our results point to an improper modulated structure where interestingly the spins oriented strictly antiferromagnetic are perpendicular to those of previously reported compounds. In the present work, a combination of magnetometry measurements, single crystal and powder neutron diffraction and density functional theory calculations have been used to accurately determine and understand the sequence of nuclear and magnetic phases present in compound **1**.

Introduction

In recent years, there has been intense interest in coordination polymers (CPs) as multifunctional materials, including dielectric and magnetic order.[1,2] Aperiodic structures are of growing interest due to the range of orders that they can facilitate. Despite the large number of known metal-organic structures,[3] CPs with aperiodicity are very scarce,[4] and magnetic structures of aperiodic CPs are almost non-existent, with the notable exception of [CH$_3$NH$_3$]Ni(HCOO)$_3$. The lack of aperiodic CPs is particularly noteworthy since weak interactions like hydrogen bonding, dipole-dipole interactions, and $\pi$-stacking, which are frequently seen in CPs, are exactly the forces that frequently cause aperiodicity in organic systems.[5] This suggests that many published CPs have non-reported modulated phases. The aperiodicity is key as it will produce distinct phonon, electronic and mass transport behaviors, and photonic properties.[5,6] Understanding aperiodic CPs is thus critical.

Hybrid organic-inorganic perovskites (HOIPs) are CPs of particular interest because of their easy fabrication, low cost, and record-high power conversion efficiency of solar cells.[7] Perovskites (ABX$_3$) consist of B-site cation which is six-fold coordinated by X-site anions, forming an octahedron. These octahedra are then arranged into a corner-sharing network with the A-site cations located in its cavities. A wide variety of functions, including dielectricity, ferroelectricity, pyroelectricity, piezoelectricity, multiferroicity, superconductivity, magnetoresistivity, and optoelectronic and electro-optic properties, can be achieved within the perovskite structure due to the chemical diversity of the A-, B- and X-site ions.[8] Moreover, neutron diffraction studies have shown that these compounds constitute excellent hosts for unusual magnetic states.[9] Carboxylate ligands, especially formate, are good mediators of magnetic interactions, so formate perovskites are of particular interest for their magnetic properties.[10]

[CH$_3$NH$_3$]Co(HCOO)$_3$ (**2**) and [CH$_3$NH$_3$]Ni(HCOO)$_3$ (**3**) are of particular interest because, uniquely among reported HOIPs, they exhibit phases with aperiodic structures (see Figure S1).[11,12] These compounds exhibit a finely balanced system driven by subtle interplay of weak interactions within the hydrogen-bond network. The formate anion serves as a hydrogen bond acceptor, while the methylammonium counterion, located within the cavities, acts as a potent hydrogen bond donor. These competing interactions result in crystal structures with nearly equivalent energies. Consequently, subtle alterations in the hydrogen bonding network can induce structural phase transitions, giving rise to modulated phases that are closely related with variations in the electronic properties of the compounds.[Error! Bookmark not defined.,13,14] **2** and **3** crystallize in the orthorhombic *Pnma* space group at room temperature. Upon cooling, at ca. 128 K and 84 K for cobalt and nickel compounds, respectively, undergo a phase transition from the orthorhombic unmodulated phase to an orthorhombic modulated phase, which crystallizes in the *Pnma*(00$\gamma$)0*s*0 space group with **q** = 0.143*c**. **2** undergoes a second phase transition below 96 K, keeping the structural incommensurability, with a significant change on the incommensurate wave-vector from **q** = 0.143*c** to **q** = 0.1247*c**. Finally, below 78 K a third phase transition takes place to a non-modulated monoclinic phase (*P*2$_1$/*n*), that remains unchanged even below the magnetic ordering temperature, $T_C$ = 16 K. On the contrary, **3** does not undergo any additional nuclear phase transitions, with its modulated structure almost invariant even below the magnetic ordering transition at $T_C$ = 34 K, which has proper magnetic incommensurability.

Some of us have recently shown that doping the B-site metals allows for the tuning of bulk magnetic properties, particularly the magnetic ordering temperature, in $[CH_3NH_3]Co_xNi_{1-x}(HCOO)_3$, with $x$ = 0.25, 0.5 and 0.75, through adjustments in metal composition.[15] Additionally, both the nuclear phase transition temperatures and their characteristics can be modulated by varying the nickel concentration stabilizing the modulated structures over a broader temperature range.

The control of the properties through the modification of the B-site metals in the formate framework inspired us to continue our investigations by investigating the isomorphous Fe(II) analogue $[CH_3NH_3]Fe(HCOO)_3$ (**1**). In this work we find that **1** also undergoes structural phase transitions to modulated structures upon cooling. We use a combination of magnetometry measurements, single crystal and powder neutron diffraction to accurately determine the sequence of phase transitions and the magnetic behavior of compound **1**.

**Experimental details**

**Sample preparation**

The synthetic route to obtain the $[CH_3NH_3]Fe(HCOO)_3$ compound is equivalent to the previously reported $[CH_3NH_3]Co(HCOO)_3$ in references **Error! Bookmark not defined.** and **Error! Bookmark not defined.**. However, $FeCl_2·6H_2O$ (3mL, 0.33 M) was used instead of $CoCl_2·6H_2O$. An additional 1.5 mL of HCOOH and 0.05 mmol of L-ascorbic acid were added to the resulting solution to be sealed in a Teflon-lined stainless steel vessel, in order to help the crystallization. This synthesis allowed us to obtain prismatic light-brown crystals of [CH$_3$NH$_3$]Fe(HCOO)$_3$ suitable for single crystal diffraction, with a yield of about 67%. The crystals were filtered, washed with ethanol (10 mL) and dried at room temperature. FT-IR (cm$^{-1}$): $\upsilon$(N–H): 3000(sh), $\upsilon_{s/as}$(N–H): 2772(w), $\delta_{as}$(N–H): 1636(w), $\upsilon$(CH3): 2873(w), 1458(w) and 1420(w), $\upsilon_{as}$(OCO): 1570(s), $\upsilon_s$(OCO): 1354(s), $\delta_{as}$(OCO): 1376(w), $\delta_s$(OCO): 796(s) and $\upsilon_{as}$(C–N): 970(w).

## Magnetic measurements

A superconducting quantum interference device (SQUID) magnetometer was used to evaluate the magnetic susceptibility of a polycrystalline sample of compound **1** in the temperature range of 2-300 K. Additionally, low field magnetic properties were investigated up to 5 T. The diamagnetism of the constituent atoms, calculated from Pascal's constants,[16] was taken into account while adjusting the experimental susceptibility. Experimental susceptibility was also corrected for the temperature-independent paramagnetism and the magnetization of the sample holder.

## Neutron powder diffraction measurements and refinement details

Neutron powder diffraction (NPD) experiments were performed on the high-flux medium-resolution diffractometer D1B at the Institut Laue Langevin (ILL, Grenoble, France) operated with a wavelength of λ = 2.521 Å (with experiment number CRG-2369). The sample (ca. 1 g) was placed in a Ø 6 mm cylindrical vanadium container inside a standard helium orange-cryostat. Two high-count diffraction patterns were recorded at 26 and 2 K, above and below the magnetic order temperature, respectively. Additionally, a thermodiffractogram in the 2-300 K range was collected in order to obtain the temperature dependent diffraction pattern of the sample. The crystal structure refinements and the magnetic structure calculations were carried out using the *FullProf Suite* program.[17]

### Single crystal neutron Laue diffraction measurements

A survey of reciprocal space as a function of temperature evolution was conducted using single-crystal neutron Laue diffraction on the CYCLOPS (Cylindrical CCD Laue Octagonal Photo Scintillator) multiple CCD diffractometer at ILL (Grenoble, France). A crystal specimen, with dimensions of 2.4 × 0.8 × 0.8 mm³, was mounted on a vanadium pin and placed in a standard orange cryostat. Multiple XYZ scans were performed to accurately center the sample on the neutron beam, optimizing the intensity of specific strong reflections. The sample was then cooled to 2 K and rotated along the φ-axis to achieve an orientation that would enable observation of magnetic peaks at low-Q values. Diffraction patterns were subsequently collected in 15-minute intervals per kelvin during warming up to 30 K. These measurements revealed the onset of magnetic order at 17 K (see Figure S2). At 30 K, the sample was reoriented to investigate a region of reciprocal space where satellite reflections were clearly visible. Laue diffraction patterns were recorded in 15-minute intervals every three kelvin between 30 and 200 K, revealing two structural phase transitions at approximately 170 K and 75 K (see Figure S2). Visualization of the Laue patterns was performed using the ESMERALDA software developed at ILL.

### Monochromatic single crystal neutron diffraction measurements

The same crystal used for neutron Laue diffraction measurements was installed on the self-dedicated low-temperature 4-circle displex on the monochromatic diffractometer D19 at ILL (Grenoble, France), and cooled with 2 K/min cooling rate. The wavelength used was 1.1698(1) Å for the RT and 1.4557(1) Å for the LT data collections, provided by a flat Cu monochromator using the 331 reflection, with $2\theta_M = 90.01°$ take-off angle, and 220 reflection with $2\theta_M = 69.91°$ take-off angle, respectively. The used wavelengths were selected based on the instrumental resolution, data completeness and to avoid overlapping of neighboring reflections in the modulated phases. The data acquisition strategy consisted on omega scans with constant steps of 0.07°, at different $\chi$ and $\varphi$ positions. In order to avoid collisions with the sample environment these omega scans covered either 79° or 64° depending on the $\chi$ angle.[18] The acquisition strategy limitation, combined with the absorption correction procedure, which rejects reflections that cannot be treated due to the material and shape of the low temperature environment, slightly reduces data completeness.

The NOMAD software developed at ILL was used for data acquisition. PFIND and DIRAX[19] programs were used to determine the unit cell, and we processed the raw data using the RETREAT program.[20] The refinements of the unit cell and offsets have been performed with the RAFD19 program. Graphical exploration of the reciprocal space of each collected data set was done using the Int3D program.[21] Possible wave-vectors were calculated at different temperatures using the Int3D and DIRAX programs. After that, each data set was indexed with a single wave-vector in the form $\mathbf{q} = \gamma \cdot c^*$, used to get a super-cell which all reflections, main and satellites, were successfully integrated. We used the SATELLITE program developed at ILL to make the decomposition into main and satellite reflections, following the super-space formalism. D19ABS program[22] was used to perform the absorption correction taking into account the experimental environment at each temperature. The structure at RT was solved by direct methods using SHELXL. Full matrix least squares refinement on $|F^2|$ using SHELXL-2014/76 as implemented in WINGX program[23] was used for structure refinement, while the for the aperiodic structures, below 170 K, the refinements were performed with the JANA2006 and JANA2020 programs. [24]

### Single crystal structural and magnetic determination and refinement details

The structure refinement of the parent orthorhombic phase (at room temperature) was performed with full matrix least squares refinement on $|F^2|$ using SHELXL[25] as implemented in WINGX program.[23] The crystal structures of the low temperature phases (150, 120, 90, 60, 45, 27 and 2 K) were solved using the SUPERFLIP program, using a charge-flipping algorithm,[26] that permitted to localize the non H-atoms. The hydrogen atoms positions were determined using the difference Fourier map in further refinement cycles. All the modulated phases, magnetic and nuclear, were refined using the super-space formalism as implemented in the JANA2006 and JANA2020 programs.[18] The (3+1)D Fourier density maps from the collected data show a modulated displacement in both framework and counterion. To model the incommensurability of the crystal structures, we include modulation waves in the position of all the atoms in the refinement. The neutron data allow us to refine all atoms with anisotropic displacement parameters (ADPs), as well as to refine the modulation in the ADPs. The *Pnma* structure refined at RT (phase **I**) transform into a modulated structure below 170 K (phase **II**). The structural modulation was refined using the *Pnma*(00$\gamma$)0*s*0 super space group with a $q_1$ = 0.1662(2)$c^*$. This wave-vector remains almost invariant in decreasing the temperature up to ca. 75 K, where compound **1** suffers a second phase transition between two modulated structures. Below 75 K (phase **III**), the former wave-vector drastically decreases, suggesting a change in the modulation period of this compound. At 60 K the refined wave-vector was $q_2$ = 0.1425(2)$c^*$ and the crystal structure was refined in the in the *Pnma*(00$\gamma$)0*s*0 super space group, as in previous phase (**II**). This modulation wave-vector remains even bellow 17 K, where the system become magnetically ordered. Below this temperature, a new phase (**IV**) is present, combining the nuclear modulation with the long range magnetic order. The magnetic signal can be indexed with a **k** = (0, 0, 0) propagation vector from the nuclear structure in *Pnma*(00$\gamma$)0*s*0 super space group. Scheme 1 shows the different phases present in compound **1** from RT to 2 K at ambient pressure.

The fact that the propagation vectors are almost commensurate allows the resolution of the structures by simply multiplying by 6 and 7 the unit cell in phases **II** and **III**, respectively.

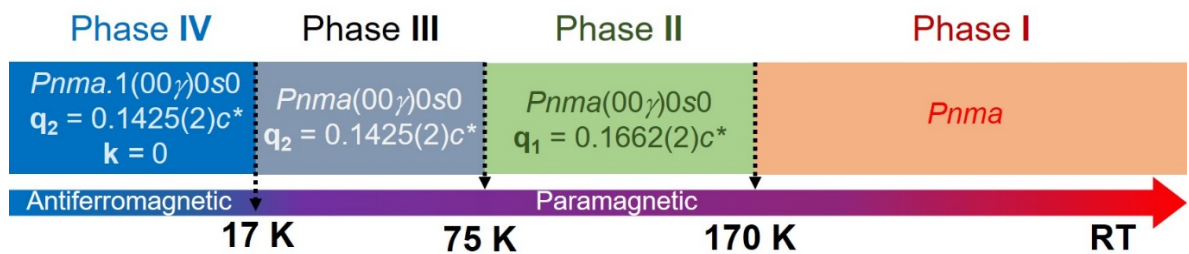

**Scheme 1.** Graphical representation of the different transition undergone by compound **1**.

However, the number of parameters to be refined is unapproachable. As an example in a previous article on compound **2**, an attempt was made to solve the modulated structure in the *P*2$_1$2$_1$2$_1$ group, with 147 atoms in the asymmetric unit, which means about 1324 parameters to be refined counting the position and the ADPs of each atom. This is why it is more convenient to solve the structure in the supergroup formalism, thus reducing the number of parameters to be refined in the structure.

**Table 1.** Crystallographic and experimental data of compound **1**, measured on the single crystal neutron diffractometer D19.

| Chemical formula | | | | $C_4H_9FeNO_6$ | | | | |
|---|---|---|---|---|---|---|---|---|
| M | | | | 223 | | | | |
| Z | | | | 4 | | | | |
| T, K | RT | 150 | 120 | 90 | 60 | 45 | 27 | 2 |
| (Super)Space group | *Pnma* | *Pnma*(00γ)0s0 | *Pnma*(00γ)0s0 | *Pnma*(00γ)0s0 | *Pnma*(00γ)0s0 | *Pnma*(00γ)0s0 | *Pnma*(00γ)0s0 | *Pnma*.1(00γ)0s0 |
| Modulation vector | - | 0.1660(2)*c*\* | 0.1663(2)*c*\* | 0.1662(2)*c*\* | 0.1425(2)*c*\* | 0.1429(2)*c*\* | 0.1425(2)*c*\* | 0.1425(2)*c*\* |
| *a*, Å | 8.5258(4) | 8.4096(7) | 8.3861(4) | 8.3861(4) | 8.3787(3) | 8.3499(2) | 8.3704(2) | 8.3708(3) |
| *b*, Å | 11.8383(9) | 11.8115(12) | 11.8084(7) | 11.8084(7) | 11.8114(6) | 11.7861(4) | 11.7988(4) | 11.7940(4) |
| *c*, Å | 8.1205(4) | 8.1449(8) | 8.1504(4) | 8.1524(4) | 8.1562(6) | 8.1500(3) | 8.1550(3) | 8.1556(4) |
| V, Å$^3$ | 819.61(8) | 809.03(13) | 807.10(7) | 807.30(7) | 807.17(8) | 802.06(4) | 805.39(4) | 805.16(6) |
| $\rho_{calc}$, mg m$^{-3}$ | 1.807 | 1.8305 | 1.8349 | 1.8345 | 1.8348 | 1.8464 | 1.8388 | 1.8393 |
| λ, Å | 1.1698(1) | 1.4557(1) | 1.4557(1) | 1.4557(1) | 1.4557(1) | 1.4557(1) | 1.4557(1) | 1.4557(1) |
| μ, mm$^{-1}$ | 0.190 | 0.215 | 0.215 | 0.215 | 0.215 | 0.215 | 0.215 | 0.215 |
| $R_1$, $I > 2\sigma(I)$ (all) | 0.0530 (0.0694) | 0.0852 (0.1068) | 0.0768 (0.1076) | 0.0929 (0.1218) | 0.0899 (0.1134) | 0.0999 (0.1174) | 0.0957 (0.1147) | 0.1031 (0.1205) |
| $wR_2$, $I > 2\sigma(I)$ (all) | 0.1022 (0.1147) | 0.1196 (0.1314) | 0.1148 (0.1599) | 0.1444 (0.1763) | 0.1361 (0.1641) | 0.1496 (0.1634) | 0.1471 (0.1617) | 0.1530 (0.1665) |
| Independent reflections | 1191 | 1334 | 3407 | 3411 | 3408 | 3494 | 3527 | 3530 |
| No. of main reflections | - | 472 | 708 | 709 | 709 | 728 | 734 | 735 |
| No. of first-order satellite reflections | - | 862 | 1268 | 1269 | 1270 | 1301 | 1318 | 1318 |
| No. of second-order satellite reflections | - | - | 1431 | 1433 | 1429 | 1465 | 1475 | 1477 |

The ratio between the number of main and satellite reflections is of 0.26 for all data sets except for that measured at 150 K, with a ratio of 0.54, since only satellites of 1st order were visible. The number of refined parameters was 277 and 107 for the paramagnetic modulated structural models, and RT, respectively. The magnetic modulated model at 2 K was refined with 278 parameters. This model was refined with the magnetic moment along $c$ direction restricted to be zero, in agreement with the magnetometry measurements. Without this restriction, convergence issues arise in the model. The modulus of the magnetic moment was constrained to be 4.0 $\mu_B$, to avoid values without physical meaning. A summary of the crystallographic and experimental data can be found on Table 1.

The DIAMOND software version 4.6.4[27] and the VESTA software version 3.4.6[28] were used to represent all phases graphically.

### Density functional theory calculations

We performed non-collinear density functional theory (DFT) calculations to probe the magnetic anisotropy of $[CH_3NH_3]Ni(HCOO)_3$ and $[CH_3NH_3]Fe(HCOO)_3$. Spin-polarized DFT

+ U method was employed in energy calculations, using the Vienna Ab initio Simulation Package (VASP).[29] In our DFT + U calculations, we used U values of 5.1 eV and 4.6 eV for $d$-electrons of $Ni^{2+}$ and $Fe^{2+}$ cations, respectively.[30] We used a plane-wave basis set with a kinetic energy cutoff of 520 eV to expand the wave functions. The Perdew-Burke-Ernzerhof functional[31] in combination with the projector augmented wave method[32] were used to solve the Kohn-Sham equations. An energy convergence threshold of $10^{-6}$ eV was used for electronic energy minimization calculations. All DFT calculations have been performed in a orthorhombic cell (4 formula units per cell), with fixed cell parameters and atomic positions of $[CH_3NH_3]Ni(HCOO)_3$[33] and $[CH_3NH_3]Fe(HCOO)_3$ (this work) determined experimentally, using a 8 × 6 × 8 k-grid (corresponding to a k-points spacing of around 0.1 Å$^{-1}$). To probe the magnetic anisotropy of both compounds, we first performed a self-consistent collinear DFT calculation. Next, a series of non-self-consistent non-collinear DFT calculations with spin-orbit coupling were performed for selected spin orientations (defined via the SAXIS keyword in VASP), after which the magnetic anisotropy of the two compounds can be obtained and compared with each other.

## Results & discussion

### Single crystal structure at RT

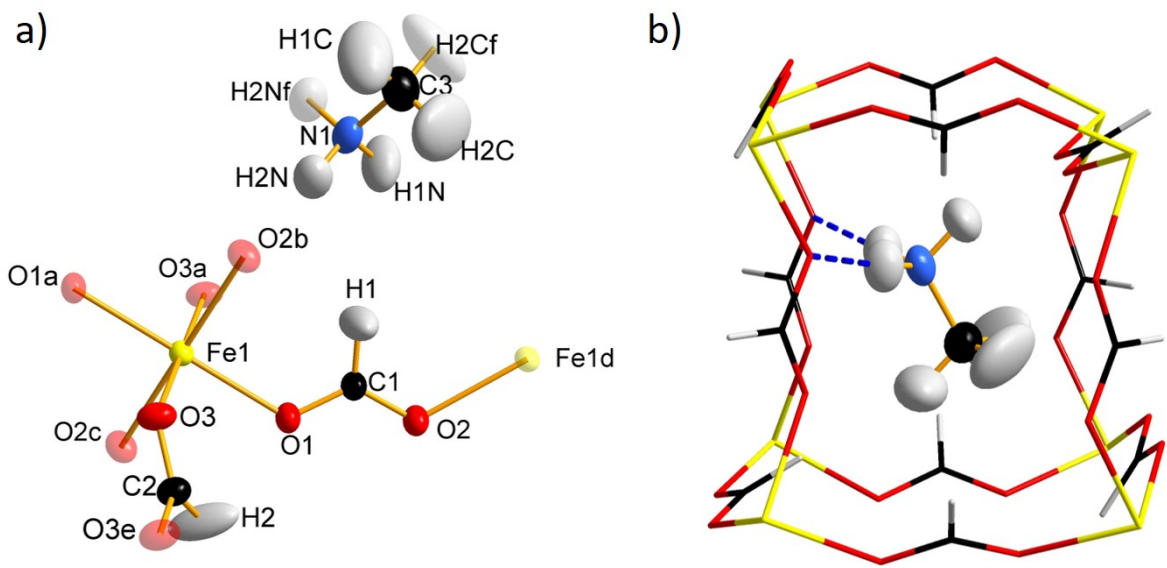

**Figure 1.** (a) View of the asymmetric unit of compound **1** where the atoms generated by symmetry are represented with transparency. (b) Representation of the methylammonium cation inside the iron-formate host framework with the possible hydrogen-bonds represented in dashed blue lines. Thermal ellipsoids at RT are calculated at 50% of probability. The irons, carbon, nitrogen, oxygen and hydrogen atoms are represented in yellow, black, blue, red and white colors, respectively. *Symmetry code*: $a = -x, -y+1, -z+1$; $b = -x+1/2, -y+1, z-1/2$; $c = x-1/2, y, -z+3/2$; $d = -x+1/2, -y+1, z+1/2$; $e = x, -y+3/2, z$; $f = x, -y+1/2, z$

The single crystal structure at RT of compound **1** has been already reported.[34] Our single crystal neutron measurements allowed us to find and refine the light atoms of the methylammonium counterion, which could not be precisely found by X-ray diffraction, in order to appropriately compare the structure with the new modulated phases. Compound **1** crystallizes in the *Pnma* space group at RT presenting a perovskite-like structure with the general formula of ABX$_3$ (phase **I**). Its structure can be depicted as a three-dimensional anionic network where each Fe(II) ion (B-sites in the perovskite notation) is surrounded by six formate ligands (X-site), which connect to other six Fe(II) ions in an *anti-anti* conformation (Figure 1). In Schläfli notation, it can alternatively be described as a 4$^{12}$6$^{3}$-pcu topology. Within this network, there is only one crystallographically independent Fe(II) atom, which sits in an inversion center. Each iron atom is surrounded by six oxygen atoms in a nearly ideal FeO$_6$ octahedron. A methylammonium counterion occupies the A-sites of the traditional perovskite model, anchored on the cavities of the anionic framework and compensating for the electronegativity of the framework, achieving the chemical neutrality. Although the empty volume of the anionic framework is 216 Å$^3$ per unit cell (26% of the total volume),[35] the methylammonium molecule does not librate randomly within the cavities as it forms two H-bonds from two of the hydrogens of the NH$_3^+$ group to two formate oxygen atoms. The third hydrogen atom does not establish any hydrogen-bond, however it is in the middle of two possible contacts. The phase transition to aperiodic structures may be caused by this structural frustration, as discussed later. Neutron single crystal diffraction allowed us to localize the H atoms and accurately refine their ADPs. From these data, we note that the hydrogen atoms of the CH$_3$- group have larger ADPs than their neighbors, indicating a possible positional disorder. This disorder has been found in the isomorphous compounds **2** and **3**, with cobalt and nickel.[11,12] Nevertheless the Fourier map of compound **1** only shows a single peak for each H atom.

### Modulated structures

The temperature evolution of the reciprocal space in compound **1** using the Laue diffractometer CYCLOPS has confirmed the number of structural phase transitions, from **I** to **IV** (see Figure S2). As is conventional, the highest temperature solid phase is denoted as phase **I**. Upon cooling, at about 170 K, the occurrence of new satellite reflections in the diffraction pattern confirm a first phase transition from phase **I** to a modulated structure determined in the *Pnma*(00$\gamma$)0*s*0 super space group, phase **II**. This transition is reminiscent of those reported for the cobalt and nickel compounds at 128 K and 84 K, respectively. However, the full data set collected at 150 K of compound **1** on the monochromatic single crystal diffractometer D19, suggests a wave vector of **q**$_1$ = 0.1662(2)*c*\*, which give almost a 6-fold increase along *c* axis, that differs from the previous reported for compounds **2** and **3** (**q** = 0.143*c*\* that corresponds with an increase of about 7-fold). The same results were obtained for the full data sets collected at 120 K and 90 K (phase **II**). While only first order satellites are visible at 150 K, second order satellites are visible when the temperature is dropped, which is related with the decrease of the Debye-Waller effect.

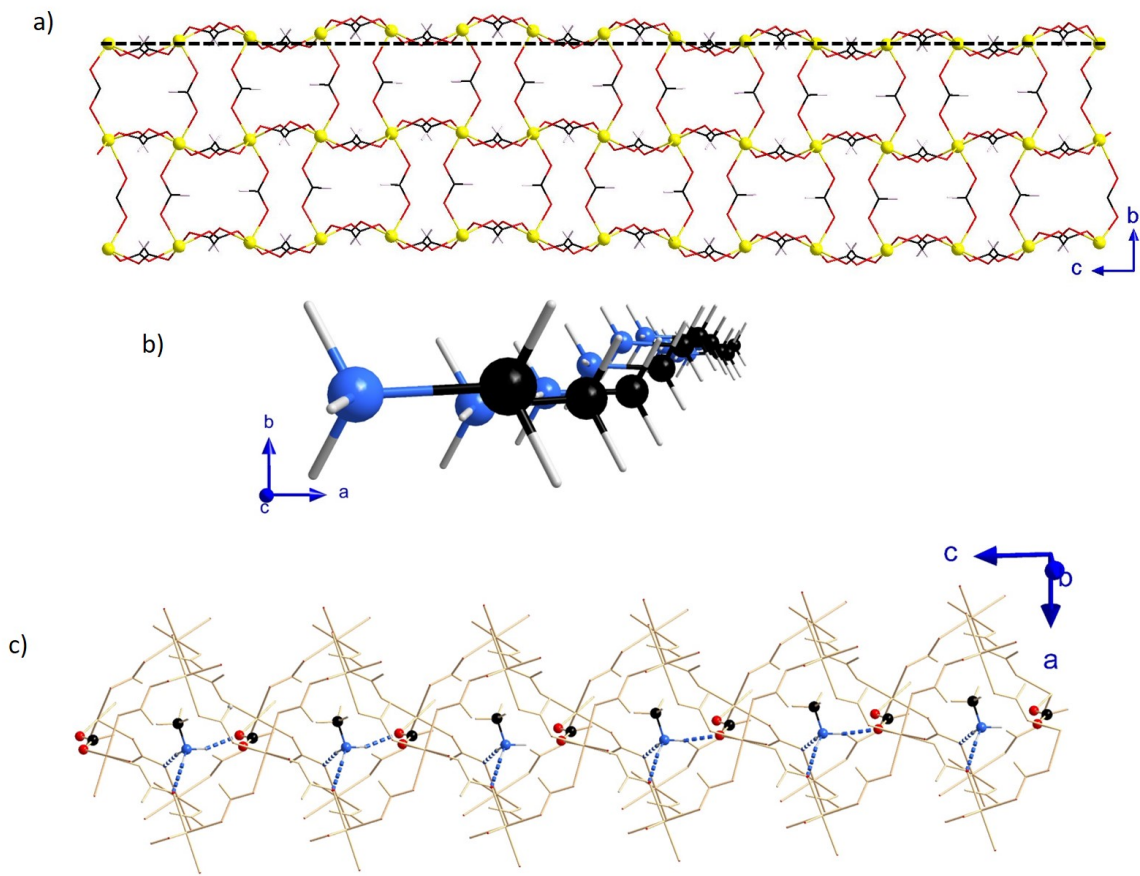

**Figure 2**. (a) Representation of the Fe(HCOO)$^-$ metal-organic framework and (b) (NH$_3$CH$_3$)$^+$ counter-ions at 90 K, showing the modulation of the atoms along the *c* axis with displacements mainly along the *b* axis. The Fe, C, O, N and H atoms are represented in yellow, black, red, blue and white colors, respectively. These graphical representations have been made taking into account a supercell (7 times the average structure along the *c* axis) to include a full period. A black dashed line is included in a) as a visual guide. (c) View of the possible hydrogen bonds involving the counterion and one of the formate ligands where it can be appreciated the flip-flop behavior of the hydrogen bond along *c* axis (contacts

H1N···O3d and H1N···O3g with $d$ = -$x$+1/2, -$y$+1, $z$+1/2 and $g$ = -$x$+1/2, $y$-1/2, $z$+1/2).
Dashed blue lines represent H-bonds.

Moreover, at about 75 K the satellite reflections become closer to the main reflections in the Laue pattern, indicating a new phase transition. The measurements on D19 at 60 K, 45 K, and 27 K indicate that this phase transition maintains invariant the super space group, but with a new wave-vector **q**$_2$ = 0.1425(2)$c^*$(phase III). This last structure remains invariant until base temperature (2 K).

In phases **II** and **III** the average structure is described using the symmetry operators of the *Pnma* space group. The application of a modulation function that exhibits sinusoidal behavior along the incommensurate parameter, *t*, changes the position of each independent atom in the average structure. Symmetry constraints are applied on the sine or cosine terms of the Fourier coefficients based on the atomic positions that are present in the *Pnma* average structure (8*d*, 4*c* or 4*b* Wyckoff positions). This implies that modulation amplitudes along the *a*, *b* and *c* axes are allowed by symmetry, thus implying small structural tilts or distortions in FeO$_6$ octahedron. Nevertheless, the amplitudes of the displacive modulation are largely across the *b* axis with mainly a little component in the *a* axis for some atoms, in both phases (see Figure 2). The refined amplitude displacements in phases **II** and **III** for Fe(II), the C and N atoms of the (CH$_3$NH$_3$)$^+$ counterion, -representing the framework and the guest molecule, respectively- are represented in Figure S3a. The amplitude displacements were shown to generally increase with cooling (see Figure S3b). Moreover, the amplitudes of the sine functions are an order of magnitude larger than the cosine functions in all atoms, which implies that the modulation in the displacement of the atoms is predominantly in phase. Despite being predominantly in phase, differences in the amplitudes of the modulation in the *b*-axis, together with small contributions from the cosine and other directions, cause the distances between the atoms to be affected by the modulation and differ across the structure, as shown in the modulation of the Fe-O bond lengths in Figure S4. These amplitude displacements for compound **1** are slightly larger than those for the previously reported compounds (see Figure 3).

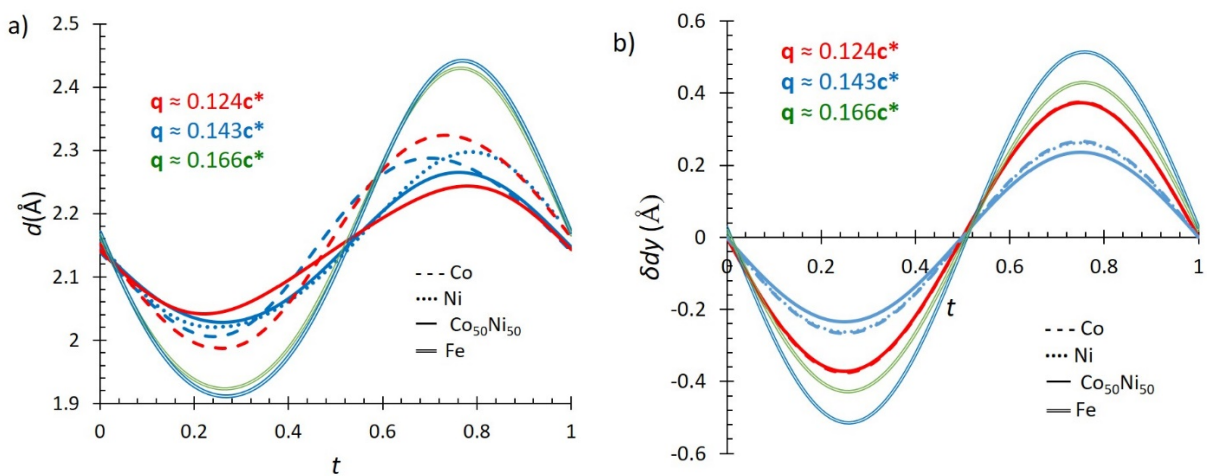

**Figure 3.** (a) H1N···O3d bond distances and (with $d$ = -$x$+1/2, -$y$+1, $z$+1/2) (b) displacement of the metal atom along *b* axis for the [CH$_3$NH$_3$]M(HCOO)$_3$ family of compounds with M = Co at 106 K (blue) and 86 K (red) (dash lines),[Error! Bookmark not defined.] M = Ni at 40 K (blue dotted lines),[Error! Bookmark not defined.] M = Co$_{0.5}$Ni$_{0.5}$ at 70 K (blue) and 30 K (red) (solid lines)[Error! Bookmark not defined.] and M = Fe at 90 K (green) and 45 K (blue) double lines. Red lines highlight the structures with the approximate wavevector **q** = 0.124$c^*$, the blue lines for the approximate wavevector **q** = 0.143$c^*$ and the green lines for the approximate wavevector **q** = 0.166$c^*$.

In phase **II** and **III** changes occur in the hydrogen-bonded network formed between the methylammonium and the oxygen atoms of the formate ligands along *c* axis, that is the direction of modulation. Like in phase **I**, the hydrogen atoms of the $CH_3$- group do not form any H bond, and two of the three hydrogen atoms of the $NH_3$- group visibly set hydrogen-bond connections towards the nearest oxygen atoms of two different formate groups. However, because of the modulation of the structure, in the aperiodic phases the last H atom of the $NH_3$- group oscillates between two oxygen atoms from the same formate ligand (Figure 2). Like in previous reported compounds, this frustration of the hydrogen-bond network must be the trigger of the phase transitions. The H1N···O3d distances (*d* = -*x*+1/2, -*y*+1, *z*+1/2) in the modulated phases **II** and **III** of compound **1** in terms of the *t* parameter are represented in Figure S3c.

In contrast to compound **2**, which loses its structural modulation at low temperature, compounds **1** and **3** maintain their nuclear modulation up to the lowest temperature studied. Therefore, the modulation of compound **1** remains invariant even below Néel temperature, where this nuclear modulation combines with the occurrence of long-range magnetic order (phase **IV**).

**Magnetic properties**

The magnetometry measurements of compound **1** reveal a global antiferromagnetic behaviour. Figure 4a depicts the temperature dependence of the susceptibility curve measured in an external magnetic field of 500 Oe. This compound under the field cooled regime, exhibits a broad peak extended from 10 to 40 K where it reaches the maximum around 17 K confirming the antiferromagnetic ground state. At low temperature, there is a plateau shape very close to zero magnetization. This indicates that in the underlying antiferromagnetic lattice the spins are coupled strictly antiparallel among them, with no signal of spin canting or non-compensated magnetic moments. In Figure 4b the isothermal magnetization at 2 K shows linearity as function of the external magnetic field, which denotes strong antiferromagnetic interactions that are not overcome even at 5 T. Neither in the magnetization temperature evolution nor in the field-dependent magnetization curves there is any indication of ferromagnetic signals in compound **1**, in contrast to compounds **2** and **3**, which show a dominant antiferromagnetic character with weak ferromagnetic spin canting.[33]

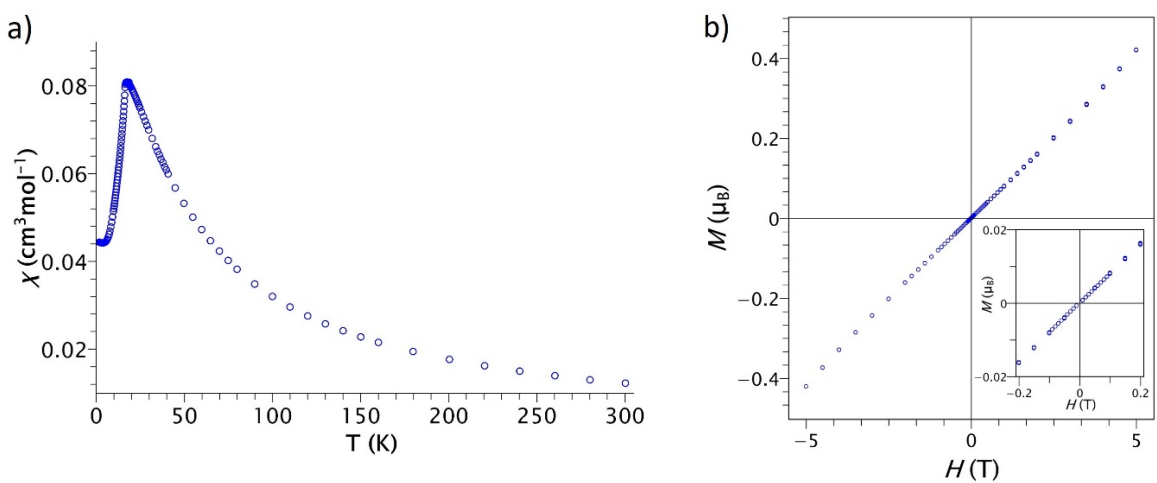

**Figure 4.** (a) Temperature dependence of the magnetic susceptibility ($\chi$) measured in a SQUID under 0.05 T from 2 to 300 K for a polycrystalline sample of compound **1**. The inset shows the inverse susceptibility in function of temperature with the Curie-Weiss fit (red line). (b) Field-dependent isothermal magnetization measured at 2 K in the range from -5.0 T to 5.0 T. Inset: close-up of data between +0.2 and -0.2 T.

**Magnetic structures**

The magnetic structure of compound **1** has been analysed through powder and single crystal neutron measurements.

**Neutron powder diffraction analysis (NPD).** Figure 5 shows the temperature evolution of compound **1**, collected on the neutron powder diffractometer D1B at ILL with $\lambda$ = 2.521 Å, from 2 to 26 K (Figure 5a) and from 26 to 300 K (Figure 5b). Only a minor change in the reflections caused by the thermal contraction is seen from RT to 26 K. When the temperature decreases, some reflections appear slightly broader; this may be due to the contribution from satellites at low temperatures. The satellite reflections are not clearly visible due to the substantial background due to the hydrogen content that produce incoherent scattering. However, there is a significant change in intensity in many main reflections at about 170 K, the temperature at which the phase transition to a modulated structure occurs (see Figure 5c). In the low temperature range from 2 to 26 K, the occurrence of new reflections is observed below 17 K, which can be indexed with a **k** = (0, 0, 0) propagation vector, using the K-search software from the FullProf distribution. The strongest magnetic reflection, corresponds to the 1 1 0 reflection, which is observed at 21.7 degrees ($2\theta$). Because we have a relatively high background in these measurements we used the difference between the measurement at 26 K and 2 K, i.e. above and below the magnetic order temperature to get the highest quality data of the magnetic diffraction (see Figure S5). This difference pattern revealed three sets of magnetic reflections, located around the $2\theta$ values of 17.5, 21.5 and 42.2 degrees. We began by analyzing the symmetry from the room temperature 'average' structure in *Pnma*, which contains only a single magnetic site (Fe(II) on the 4*b* Wyckoff position). Representational analysis from this *Pnma* model with the propagation vector **k** = (0, 0, 0), produces four irreducible representations corresponding to the four Shubnikov magnetic groups; *Pnm'a'*, *Pnma*.1, *Pn'm'a* and *Pn'ma'*. The magnetometry measurements of compound **1** shows a global antiferromagnetic behavior, hence models in each of the four magnetic space groups are potential solutions. We therefore conducted fits of the four models to the data in order to distinguish between them.

The refinement performed on the magnetic space group *Pnm'a'* is not able to correctly fit all the magnetic reflections with only the 1 0 0 reflection fitting well. Therefore, it can be discarded. The *Pn'm'a* and *Pn'ma'* magnetic groups allow by symmetry ferromagnetic components along the *c*- and *b*-axes, respectively. However, if these components are fixed to zero both models may be considered as a possible magnetic space group. The *Pn'm'a* Shubnikov group fit results, after fixing the component along the *c* axis to zero, in a pure antiferromagnetic structure with the main component of the magnetic moments coupled antiferromagnetically along the *b* axis. A weak canting along the *a* axis is present, which is also antiferromagnetically coupled along this axis. However, the $R_f$ > 20%, which is slightly larger than maybe expected. The magnetic space group *Pn'ma'*, gives a good fit with a low $R_f$ factor of about 15 %, though the reflections with calculated zero intensities do not contribute to the R-factor. However, this model does not fit properly the first set of magnetic reflections, 0 0 1 and 1 0 0 (see Figure S6). Additionally, the value of the refined magnetic moments, ca. 3.6(1) $\mu_B$, is slightly lower than expected for a Fe(II) compound. The last magnetic model has the magnetic space group *Pnma*.1 (see Figure S6), which is also the only magnetic space group that does not allow a net ferromagnetic component and is guaranteed by symmetry to be an antiferromagnetic structure. The refinement was initially unstable due to convergence problems in the component along the *c* axis of the magnetic moment. The most sensitive reflection to the $M_z$ component of the magnetic moment is the 1 1 1 reflection; this reflection is structurally a forbidden reflection but is allowed by symmetry in the magnetic space group. From the difference pattern no intensity is observed in this reflection suggesting that the component along the *c*-axis of the magnetic moment is zero within the accuracy of the current measurements. The refined magnetic structure, with the component along the *c* axis constrained to zero, is characterized by magnetic moments coupled strictly antiferromagnetic in the *ac* plane through *M*-OCO-*M* pathways in [1 0 1] and [-1 0 1] directions. The moments primarily point along the *a* direction, with a minor component along the *b* direction. These layers are stacked along the *b* axis in an ABAB sequence, where the $M_y$ component retains its orientation, while the $M_x$ component alternates between planes (see Figure 6). This model reproduces well the full set of observations, with $R_f$ = 16 % that is close to that obtained with the *Pn'ma'* magnetic space group. The refined value of the magnetic moment, 4.2(1) μ$_B$, corresponds within the error with the theoretical value for a high-spin Fe(II)-based compound (S = 2). Therefore, we consider that the *Pnma*.1 is the most plausible average magnetic space group for this compound.

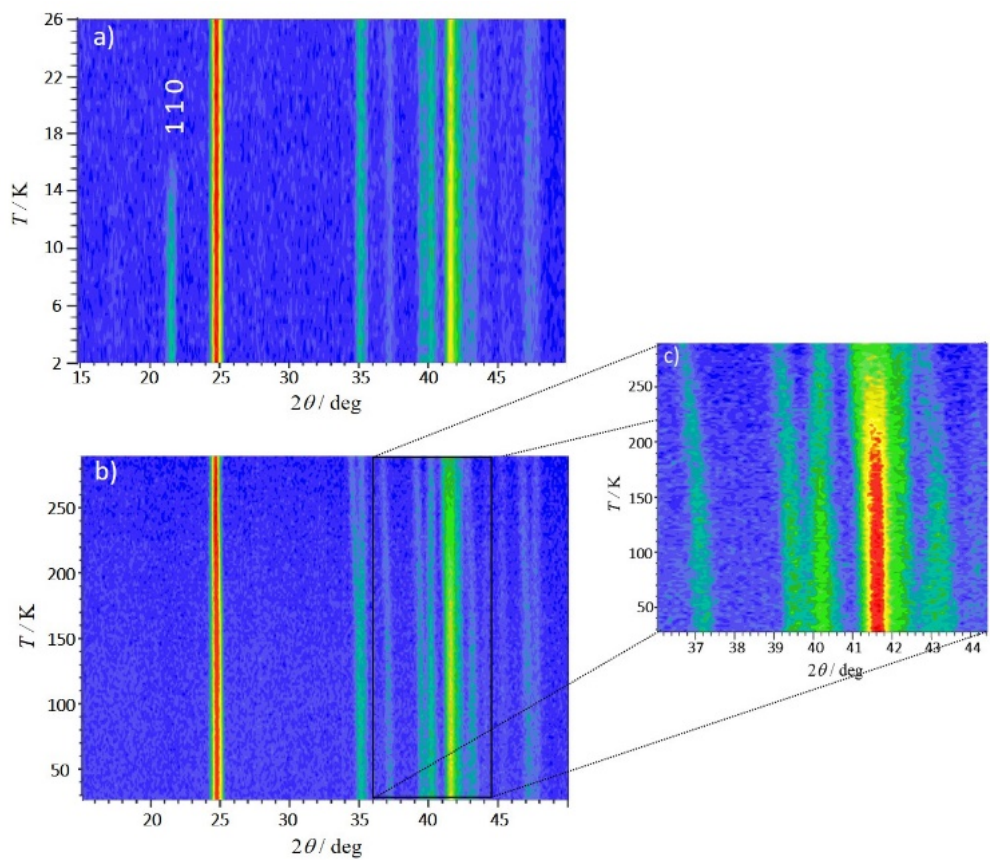

**Figure 5**. Mesh plots of compound **1** with the intensities represented in logarithmic scale, of the thermodiffractograms collected at D1B with λ = 2.521 Å. (a) Measurement in the temperature range of 2-26 K. The appearance of new reflections and the increase of intensity on the top of the nuclear reflections below 17 K, is in agreement with the magnetometry measurements. (b) Measurement in the temperature range of 26-300 K. The figure shows

the changes of intensities of the nuclear reflections from RT to the phase transition temperature at *ca.* 200 K. Below this temperature the orthorhombic phase become modulated. (c) Measurement in $2\theta$ range from 36° and 44.5° between 26 K and 300 K, showing a significant the change in intensity of the 0 3 1 and 2 1 1 sets of reflections at 41.7° and 41.4°, respectively, at about 170 K, where the transition between phase **I** and **II** occurs.

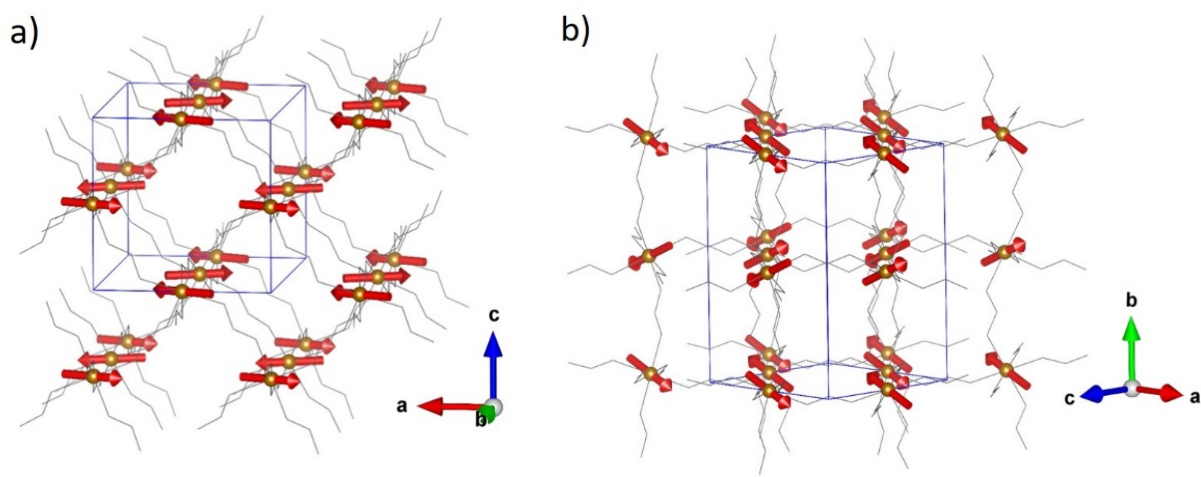

**Figure 6**. Perspective view along the *b* axis and [1 0 1] direction of the magnetic structure refined in the *Pnma*.1 Shubnikov space group for compound **1** (left and right, respectively) from NPD measurements. The nuclear and magnetic unit cell [**k** = (0, 0, 0)], has been represented in blue and the magnetic moments in red arrows. For clarity, the counterion and hydrogen atoms have been removed, and *M*-OCO-*M* pathways has been represented as grey sticks.

This is a different magnetic structure from both **2**, which exhibits a non-collinear antiferromagnetic structure where magnetic moments are not strictly compensated in the orthorhombic *b* axis (*Pn'ma'* and *P2'₁/n*)[13] and **3** which presents a similar magnetic structure to that of **2** but with the modulation of the magnetic moments along *c* axis (*Pn'ma'*(00γ)0s0).[12] In these structures, the magnetic moments lie principally on the *ac* plane with a small component in *b*. Within the *ac* plane, each moment is coupled to its nearest neighbor through *M*-OCO-*M* superexchange pathways, resulting in ferromagnetic alignment of the $M_x$ and $M_y$ components, while $M_z$ is antiferromagnetically coupled. Along the *b* axis moments are also coupled via *M*-OCO-*M* pathways, with $M_x$ and $M_z$ components coupled antiferromagnetically while $M_y$ component is aligned ferromagnetically. These models present a weak ferromagnetic correlation caused by the canting of the moments along the *b* axis of about 0.68(12) $\mu_B$ and 0.55(9) $\mu_B$ for **2** and **3**, respectively. This change in the magnetic structure is entirely consistent with the notable difference observed in the magnetometry measurements: *Pnma*.1 forbids any net moment, whereas *Pn'ma'* permits a net moment. The unusually large R-factors are caused by the lower signal to noise ratio in the diffraction data, however the symmetry analysis allows us to be confident in the general features of the structure.

**Single crystal neutron diffraction analysis.** Nevertheless, to improve the quality of our model we then carried out single crystal neutron diffraction on large single crystals to refine our magnetic structure further. Single crystal diffraction minimizes the effect of the hydrogen incoherent background, which was significant in our neutron powder diffraction data. This is because single crystal diffraction measures only a small solid angle around each reflection, unlike in powder diffraction, where a large volume of reciprocal space contribute to each data point, and so the isotropically incoherent hydrogen scattering is much less significant.

We measured a full data set on the D19 diffractometer at the ILL at 2 K. In this case the higher quality of the data allowed us to carry out symmetry analysis using two independent wave-vectors, the first one associated to the structural distortion, $q_2 = 0.1425(2)c^*$ from phase III, and the second one related to the magnetic contribution. As the magnetic order occurs in a nuclear structure that is already modulated, the magnetic propagation vector must be calculated considering the super space group and not only the average structure. The examination of the D19 diffraction pattern at 2 K (phase **IV**) show that there are no extra reflections compared with the 27 K pattern (phase **III**), then all the magnetic signals are on top of structural reflections, main- or satellite-reflections, which suggest a $k = (0, 0, 0)$ magnetic wave-vector, in agreement with the NPD results. Taken together, the magnetic wave-vector contributes to the nuclear modulated structure and the magnetic super-space group cannot contain the operator $\{1'|0001/2\}$.[36] The analysis using ISODISTORT[37] with $q_2 = 0.1425(2)c^*$, as a displacive, and $k = (0, 0, 0)$, as a magnetic, wave vectors give us a list of 32 possible super space magnetic groups (see Table S1). However, this list can be substantially reduced taking into account the observed antiferromagnetic order in magnetometry measurements and NPD results. The *Pnma*.1 magnetic space group provides the best results of the neutron powder data, considering the average structure, as was discussed in the preceding section. Furthermore, the paramagnetic nuclear structure at 27 K crystallizes in the *Pnma*(00$\gamma$)0s0 super space group, then the most probable magnetic super space group is the *Pnma*.1(00γ)0s0, that is number 62.1.9.2.m441.1.

Considering this magnetic super space group, we then focused on analyzing the different distortion modes of the Fe(II) atoms. As the unit cell parameters were fixed in indexing, it was not necessary to refine any modes corresponding to global strain. With this consideration, the *Pnma*.1(00γ)0s0 magnetic super space group give us nine distortion modes for the Fe(II) magnetic atom: three displacive modes (along *a*, *b* and *c*) which are responsible of the structural modulation and correspond to the sine terms shown in Figure S3a; and six modes, which are purely magnetic. These are the *x*, *y*, and *z* components of the average moment, which are expected to be very similar to the average magnetic structure found for the powder diffraction data using just $\mathbf{k}$ = (0, 0, 0), and three sinusoidal modulations with amplitudes along *x*, *y*, and *z*, which allow for an incommensurate modulation of the moment direction and magnitude. Only these last three "sinusoidal" modes contribute to magnetic incommensurability, with the other six modes either being purely structural or only altering the magnetic average structure. We find that refining these sinusoidal proper magnetic modes responsible of the proper magnetic modulations does not improve significantly the goodness of fit and so thus, we conclude that these three sinusoidal modes are not activated in this structure. This is different to compound **3**, where the sinusoidal proper magnetic modes were not negligible, leading to proper magnetic incommensurability in this compound. Additionally if we refine freely the modulus of the magnetic moment of the iron(II) atoms, we obtain a value of $|M_{Fe(II)}|$ = 4.256(33) $\mu_B$, which is larger than the expected for a high spin Fe(II) compound (ca. 4 $\mu_B$) ), and suggests the presence of significant orbital moment, as anticipated for the $^5$T term. It is important to note that this value is in agreement with the obtained result of the NPD [4.2(1) $\mu_B$] using the average structure. In the final cycles of refinement of the single crystal data the modulus of the magnetic moment of the Fe(II) atoms was restricted to be 4 $\mu_B$ (corresponding to a S = 2) in order to consider only the spin contribution. Considering the complete structure, 278 parameters were refined, 277 to refine the structural modulation using up to second harmonic modulation waves and only one to refine the magnetic structure, using 3530 independent reflections of which 2512 are strong reflections with $I > 3\sigma(I)$. The modulated magnetic structure is essentially the same as the magnetic structure obtained with NPD except for the modulation in the position of the irons. The magnetic moments of the iron atoms are coupled strictly antiferromagnetic through *M*-OCO-*M* pathways in [1 0 1] and [-1 0 1] directions forming antiferromagnetic layers in the *ac* plane. The magnetic moments are contained in the *ab* plane, pointing mainly along the [1 1 0] direction (see Figure 7). Moreover, these AF layers are piled along the *b* direction in an ABAB sequence, where the $M_y$ component retains its orientation, while the $M_x$ component alternates between planes.

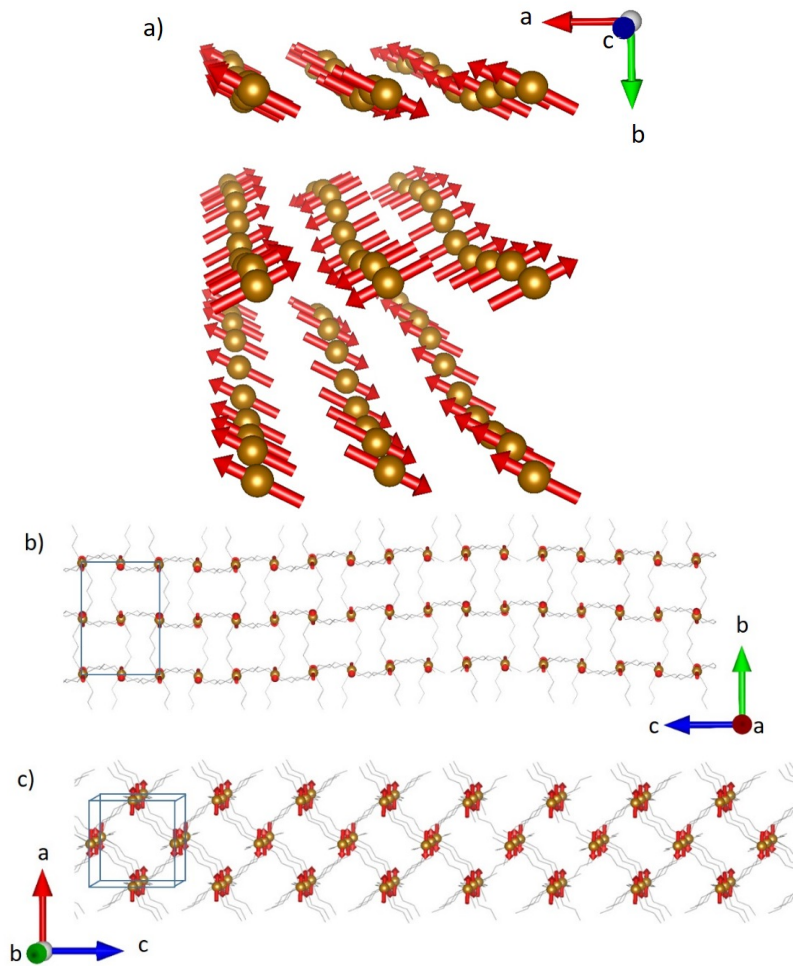

**Figure 7.** View of the refined magnetic moments (red arrows) of Fe(II) in the *Pnma*.1(00γ)0s0 magnetic super space group along the *c*, *a* and *b* directions, a), b) and c), respectively. The graphical representation was carried out taking into account a supercell that is 8 times the average structure along the *c* axis in order to take into account at least one full period. The average unit cell has been represented in blue and the magnetic moments in red arrows. For clarity, only the Fe(II) atoms (brown color) have been shown in a), and the counterion and hydrogen atoms have been removed, and *M*-OCO-*M* pathways has been represented as grey sticks in b) and c).

**Theoretical calculations**

The refined magnetic structure obtained for compound **1** is substantially different from those of compounds **2** and **3**, although their nuclear structures are broadly similar, with identical tilting patterns and similar methylammonium orientations within the cage.[12,13] The previous reported magnetic structures present their magnetic moments pointing principally along the modulated direction, that is the *c* axis, in contrast with the iron(II) magnetic moments that are rigorously perpendicular to that direction. To get a better understanding on the difference in spin orientation between nickel and iron compounds, we performed non-collinear DFT/GGA+U calculations to examine the magnetic anisotropy of Fe and Ni formate perovskites, based on their experimentally determined orthorhombic structures. Initially, a self-consistent collinear calculation was performed for a configuration with G-type antiferromagnetic (AFM) spin ordering. Subsequently, magnetic anisotropy was assessed through a non-self-consistent approach by utilizing the wavefunction and charge densities from the collinear AFM calculation and orienting the spins along the [0 0 1], [0 1 0], [1 0 0], [1 1 0], [1 0 1], and [0 1 1] axes by assigning different SAXIS values. The results show significant magnetic anisotropy in the Fe system (see Table 2), whereas the Ni system does not exhibit noticeable magnetic anisotropy within the uncertainty limit with our current computational settings.

**Table 2.** Relative energies (in meV per $Fe^{2+}$) of a magnetic configuration of [CH$_3$NH$_3$]Fe(HCOO)$_3$ along different spin directions obtained through non-self-consistent non-collinear DFT/GGA+U calculations. Note the spin directions are along the simulation cell, not along the atomic positions.

| Direction | ΔE (meV per $Fe^{2+}$) |
|---|---|
| [0 0 1] | 0.37 |
| [0 1 0] | 0.43 |
| [1 0 0] | 0.00 |
| [1 1 0] | 0.22 |
| [1 0 1] | 0.19 |
| [0 1 1] | 0.40 |

### Discussion

Crystal structure analyses of compound **1** through neutron diffraction have revealed two unreported nuclear phase transitions at 170 K and 75 K involving modulated structures. The first one implies a transition from the phase **I** in the *Pnma* orthorhombic space group to the phase **II** described in the *Pnma*(00$\gamma$)0*s*0 super space group with a wave-vector $q_1 = 0.1662(2)c^*$, which give almost a 6-fold increase along the *c* axis. In the second transition the satellite reflections become closer to the main reflections, which implies a change of wave-vector to $q_2 = 0.1425(2)c^*$ (phase **III**), that corresponds to a 7-fold increase along the *c* axis, keeping invariant the super space group. This series of phase transitions is reminiscent of those undergone by the isomorphous reported cobalt-based (**2**) and nickel-based (**3**) compounds (see Scheme 2), as well as the solid solutions [CH$_3$NH$_3$]Co$_x$Ni$_{1-x}$(HCOO)$_3$, with $x = 0.25, 0.5$ and $0.75$. Compound **2** suffers the first periodic to incommensurate phase transition at lower temperature (128 K) being the wave vector of the incommensurate phase $\mathbf{q} = 0.143(2)c^*$ (7-fold). The second phase transition at 96 K, involves a different wave vector $\mathbf{q} = 0.1247c^*$, corresponding to a 8-fold increase along the $c$ axis. Moreover, it undergoes a third phase transition to a monoclinic $P2_1/c$ space group, where the system became twined with two main domains with a relation between them of 180° around the monoclinic $a^*$ axis [8.1621(3) 8.2487(3) 11.6584(4) 90 91.891(3) 90], that corresponds with the $c^*$ axis in the orthorhombic setting. From the analysis of the phase transitions of previously published compounds, it could be seen that the wave-vector becomes smaller with the decrease of temperature being as consequence the modulation period longer. Compound **3** only present one phase transition from the *Pnma* space group at RT to the *Pnma*(00$\gamma$)0$s$0 super space group with a wave vector $\mathbf{q} = 0.143c^*$ (7-fold) at 84 K. Similar to compound **1**, compound **3** maintain the modulated structure below the magnetically ordered temperature. In the modulated structures not only the position of the guest molecules (methylammonium cation) is modulated, but also the positions of the atoms in the anionic 3-dimmensional framework. The modulation of the structure is stablished along the $c$ axis, as it reflects the $\mathbf{q}$ vectors, and the amplitude of the modulation appears mainly along the $b$ axis with a small component along the $a$ axis, for all the aperiodic phases. It can be seen from our results that the iron compound shows higher modulation amplitudes than those found in the previously published cobalt and nickel compounds (see Figure 3b). The analysis of the hydrogen network suggests that the trigger of the phase transitions involving modulated phases must lie on the frustration existent between the two possible contacts of one of the hydrogen atoms of the NH$_3$- group of the methylammonium cation, and two oxygen atoms from the same formate ligand (H1N towards O3d and O3g with $d$ = -$x$+1/2, -$y$+1, $z$+1/2 and $g$ = -$x$+1/2, $y$-1/2, $z$+1/2). Because of the modulation of the position of the methylammonium together with the atomic positions of the host framework, these two interactions evolve along the $c$ axis (see Figure 3a). This aligns with prior studies, which suggest that hydrogen bonding exhibits approximately three times greater strength in formate perovskites compared to other hybrid perovskites.[38] Figure 3a also shows that the increased modulation amplitude in compound **1** influences H-bond contacts, resulting in greater distance variations compared to previously reported structures.

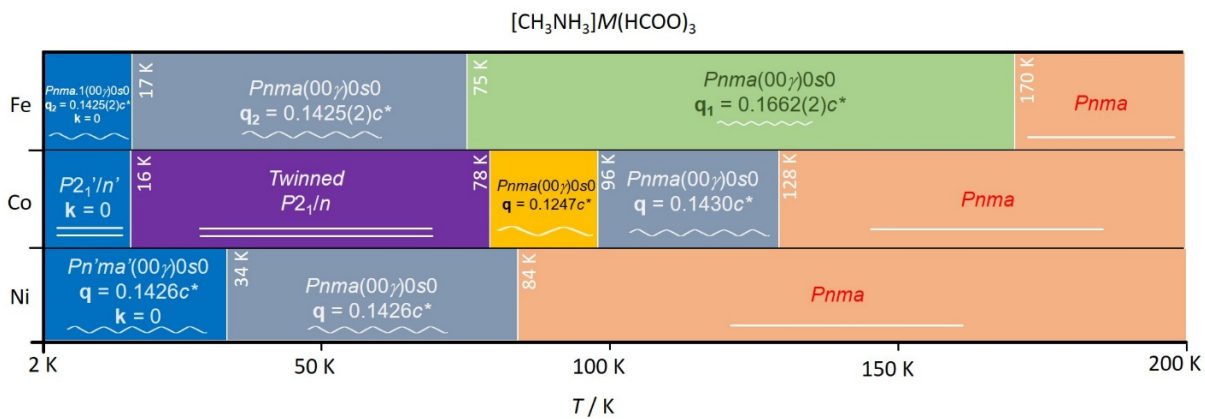

**Scheme 2.** Graphical representation of the different transition undergone by compounds **1**, **2** and **3**.

All compounds present long-range magnetic order at low temperatures. In compounds **1** and **3** the long range magnetic order coexist with nuclear modulated structures below 17 K and 34 K, respectively. In compound **3** coexistence of structural and proper magnetic incommensurability was observed, since the proper magnetic "sinusoidal" modes were non-negligible. In other words, the magnetic moments of nickel atoms vary as a function of the *t* parameter in modulus and orientation, rather than simply following structural displacement. The magnetic moments of the nickel atoms are oriented mainly along *c* direction, with a small tilt along *b* direction that varies in function of *t*. The magnetic modulation amplitudes act along the *a* axis, that is perpendicularly to the main structural modulation amplitude. The magnetic moments are ferromagnetically coupled along the *c* direction, and antiferromagnetically coupled along *a* and *b* directions. Compound **1** presents quite different magnetic structure. The magnetic moments of the iron atoms are contained in the *ab* plane, pointing mainly along the [1 1 0] direction. Although the magnetic moments are coupled in the same way as compound **3**, that is ferromagnetically along *a* and *b* and antiferromagnetically along *c*, the orientation of the moments are perpendicular to those of previous reported compounds. Moreover, the magnetic moments in compound **1** do not differ neither in modulus nor in orientation in function of *t*. This lack of modulation in the magnetic moments prevents a magnetic incommensurability with proper character.

Recent studies have revealed an intriguing behavior of compound **3** when magnetization measurements are made at 2 K.[39] Interestingly, the first magnetization of this compound takes a different route than successive magnetizations when the hysteresis loops of this compound are analyzed. The most probable explanation is that some of the active magnetic modes were disabled due to the influence of an applied external field greater than 500 Oe. The resulting magnetic structure, a magnetically modulated collinear structure, is energetically more favorable, which accounts for the increase in the value of the coercive field. The suppression of the proper magnetic modes slightly modifies the magnetization values, which explains why the magnetization cycles have no other feature than a change in sign at the critical field. Moreover, the $[CH_3NH_3]Co_{0.5}Ni_{0.5}(HCOO)_3$ solid solution presents magnetic order below 22.5 K also with coexistence of the modulated structure. Its magnetic structure is similar to that of compound **3**, however the presence of Co(II) in the metal site precludes the modulation of the magnetic spins, giving rise to an improper magnetic structure. These structures without proper modulation resembles the magnetic structure of compound **1**. Although proper modulation in magnetic structures is possible strictly antiferromagnetic (compound **1**) and spin canted (nickel containing compounds) structures, the activation of these modes makes the final magnetic structure energetically less favorable.

Non-collinear DFT calculations based on the experimentally determined structures of compounds **1** and **3** indicate that, while the [1 0 0] direction is energetically the most favorable in compound **1** (see Table 2), compound **3** does not show noticeable energy difference along different spin directions. This suggests that the Fe system exhibits significant magnetic anisotropy, whereas the Ni system does not, resulting in distinct spin orientations in the two compounds. The perpendicular switch in spin direction in function of the M(II) ion has also been observed in other formate families of compounds. For example the $NH_4M(HCOO)_3$ (with M = $Mn^{2+}$, $Fe^{2+}$, $Co^{2+}$ and $Ni^{2+}$) series of compounds, that crystallize in the $P6_322$ space group at RT.[40,41] In these compounds, the long-range magnetic order produce antiferromagnetic structures where the spin directions were refined to be in the *ab*-plane for manganese- and cobalt-based compounds (with the $P2_1.1$ or $P2_1$' magnetic space groups, respectively) and along the *c* axis for the iron- and nickel-based compounds (with the $P6_3$' and $P6_3$'22' magnetic space groups, respectively).[42] Another example is the mixed valence $[(CH_3)_2NH_2]M^{II}Fe^{III}(HCOO)_6$ niccolite-like compounds. These compounds crystallize in the *P*-31*c* space group at RT, where the amine group rotates inside the cavities of the host formate framework. At ca. 155 K the M(II) = $Fe^{2+}$ compound presents a structural phase transition to *R*-3*c* space group, because of the block of the amine group in three different positions along the *c* axis. Below $T_N$, compounds with M(II) = $Mn^{2+}$ and $Co^{2+}$ present long-range antiferromagnetic order in the *C*2'/*c*' magnetic space group, with their spins mainly contained in the *ab*-plane.[43] However, the magnetic moments in the iron-based compound, which present a magnetic structure in the *R*-3*c*' magnetic space group, prefers to align along the *c* axis.[44,45] This change in spin orientation as a function of divalent metal correlates with the shift in direction of the anisotropy easy axis for the different M(II) cations.

## Conclusions

Neutron diffraction analysis of the crystal structure of compound **1** identified two previously unreported nuclear phase transitions occurring at 170 K and 75 K. These transitions involve modulated structures, with the first marking a shift from the *Pnma* orthorhombic space group to the *Pnma*(00γ)0s0 super space group, leading to an almost six-fold increase along the *c* axis. This **q**-vector has not been observed in previous structural analyses of this family of compounds. The second transition is characterized by a change in the wave-vector, resulting in a seven-fold increase along the *c* axis, while the super space group remains unchanged. The observed modulation in these phases is driven by the hydrogen bond network within the structure. In formate perovskites, hydrogen bonds are notably stronger than in other hybrid perovskites, which contributes to the complex structural modulation.[38] These phase transitions are similar to those observed in related cobalt- and nickel-based compounds, where modulation periods become longer with decreasing temperature. It could be observed that the modulate structures appear at higher temperature using less electronegative metal atoms, being 84 K, 128 K and 170 K for Ni, Co and Fe, respectively, which give us a clue in the design of modulate structures. Furthermore, previous analysis of the solid solutions [CH$_3$NH$_3$]Co$_x$Ni$_{x-1}$(HCOO)$_3$ with *x* =0.25, 0.5 and 0.75 shown that the mixing of different B-site metal atoms increases the frustration in the structure, which stabilizes modulated structures over a broader temperature range. These results advocate that doping a structure with less electronegative metals could give rise to stable modulated structures at higher temperatures and over a wider range of temperatures.

Magnetically, compounds **1** and **3** demonstrate long-range magnetic order at low temperatures, coexisting with nuclear modulated structures. For compound **3**, both structural and magnetic incommensurability were observed, with magnetic moments modulated sinusoidally. In contrast, compound **1** displays a magnetic structure confined to the *ab* plane, differing from other compounds, with magnetic moments that do not exhibit modulation. The absence of modulation in the magnetic structure of compound **1** is reminiscent of the behavior of compound **3** under an external magnetic field or that of the [CH$_3$NH$_3$]Co$_{0.5}$Ni$_{0.5}$(HCOO)$_3$ solid solution. This suggests that, while magnetic modulation is possible in all compounds, it is not energetically favorable, which implies that external stimuli or changes at the metal site prevent proper magnetic incommensurability. Moreover, non-collinear DFT calculations reveal that the iron system exhibits significant magnetic anisotropy, while the nickel system does not. This difference in magnetic anisotropy may explain the distinct spin orientations in compounds **1** and **3** as observed in our experiments.

With this study, we aim to clarify the mechanisms behind transitions involving modulated nuclear structures and the formation of both proper modulated magnetic structures, where magnetic moments vary with the parameter *t*, and improper modulated structures, where they do not. The investigation of modulated structures is crucial for enhancing our understanding of the structure-property relationships in coordination polymers (CPs). While aperiodic molecular frameworks are still rarely reported, this work offers valuable insights into the interactions driving these phases, opening opportunities for the rational design of molecular compounds with modulated phases and more precise control over their properties.

**Author contributions**

L.C.D. and O.F. carried on the conceptualization of the work; L.M. synthesized the samples; O.F. carried out the magnetic measurements; L.C.D. carried out the single crystal neutron diffraction experiments and data analysis; O.F. and L.C.D. carried out the powder neutron diffraction experiments and data analysis; S.L. carried out the DFT calculations; L.C.D. wrote the original draft; M.J.C., O.F. and S.L. carried out the review and editing of the article.

**Corresponding Author**

\* Laura Cañadillas-Delgado, *Institut Laue-Langevin, 71 Avenue des Martyrs, CS 20156, 38042 Grenoble Cedex 9, France. E*mail: lcd@ill.fr

**Data availability**

Crystallographic data, in CIF format, for the structures of **1** collected at 150, 120 and 90 K (phase **II**), and 60, 45 and 27 K (phase **III**) have been deposited at The Bilbao Incommensurate Crystal Structure Database with entry codes BlnUw5XO0aQ and 4ZvEBIyCMDd, for phase **II** and **III** respectively. The crystallographic data, in CIF format, for the commensurate structures at RT of **1** can be downloaded from the Cambridge Crystallographic Data Centre through the CCDC reference number 2406391. Data for this article, including single crystal neutron diffraction measurements, are available at ILL-DATA 2406391at https://doi.ill.fr/10.5291/ILL-DATA.5-12-339. The software for refinement of modulated structures JANA2006 and JANA2020 can be found at http://jana.fzu.cz/. The version of the code employed for this study are version JANA2006 and JANA2020. The software for powder data treatment FullProf Suite can be found at <https://www.ill.eu/sites/fullprof/php/downloads.html>. The version of the code employed for this study is FullProf_Suite Windows(64 bits). The software for refinement of the non-modulated structure WinGX can be found at https://www.chem.gla.ac.uk/~louis/software/wingx/. The version of the code employed for this study is version 2014.1. The software for DFT calculations Vienna Ab initio Simulation Package (VASP) can be found at https://www.vasp.at/.

**Conflicts of interest**

There are no conflicts to declare.

ACKNOWLEDGMENT

We are grateful to the Institut Laue Langevin and Spanish-CRG instrument D1B for the allocated neutron beam-time (Proposals No. CRG-2369 on D1B and 5-12-339 on D19). S. L. acknowledges the use of the Sulis supercomputer through the HPC Midlands+ Consortium and the ARCHER2 supercomputer through membership of the UK's HPC Materials Chemistry Consortium, which are funded by EPSRC Grants EP/T022108/1 and EP/X035859/1, respectively

Metal-Organic Compound [NH$_2$(CH$_3$)$_2$][Fe$^{III}$Fe$^{II}$(HCOO)$_6$]. *IUCrJ* **2020**, *7* (5), 803–813. https://doi.org/10.1107/S205225252000737X.

# Supplementary information

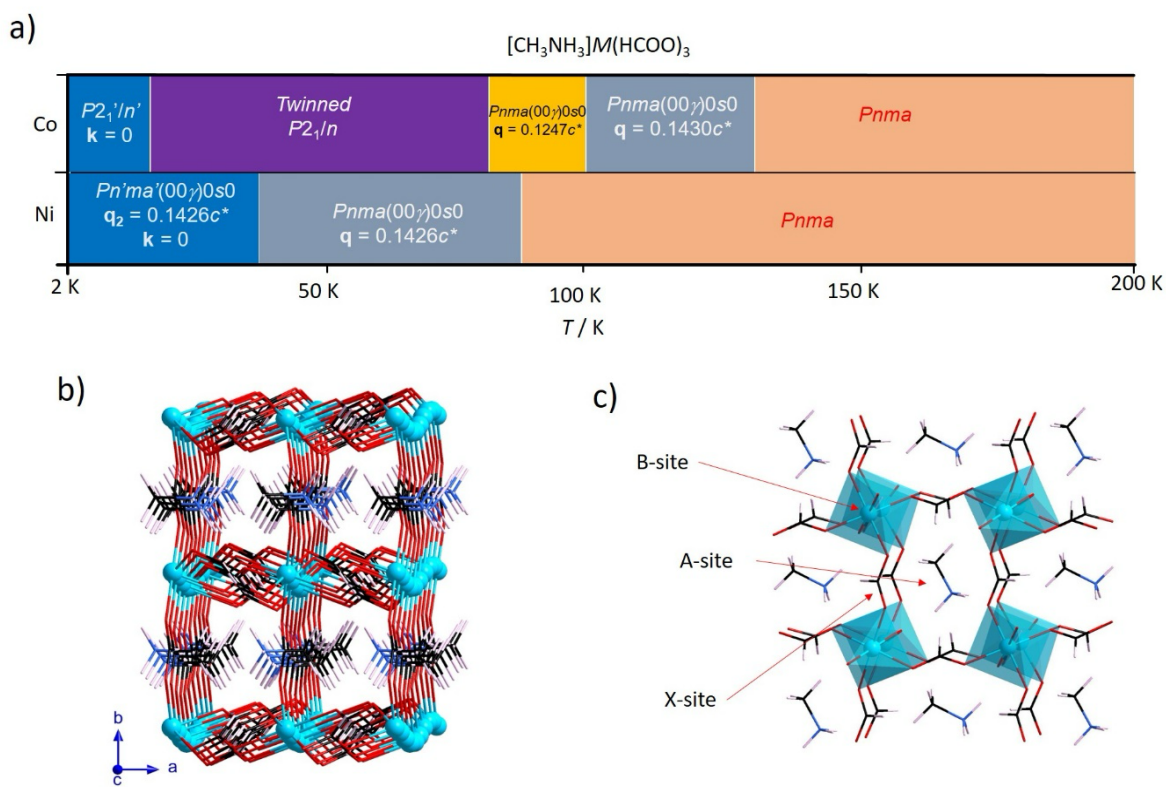

**Figure S1**. (**a**) Graphical representation of the different transition undergone by compounds **2** and **3**. (**b**) View of the modulated three-dimensional structure of compounds **2** and **3**, where it could be appreciated the displacement of the atoms mainly along the *b*-axis. (**c**) View of the Perovskite-like framework. Carbon, oxygen, hydrogen, nitrogen and metal atoms are represented in black, red, pink, blue and light blue, respectively.

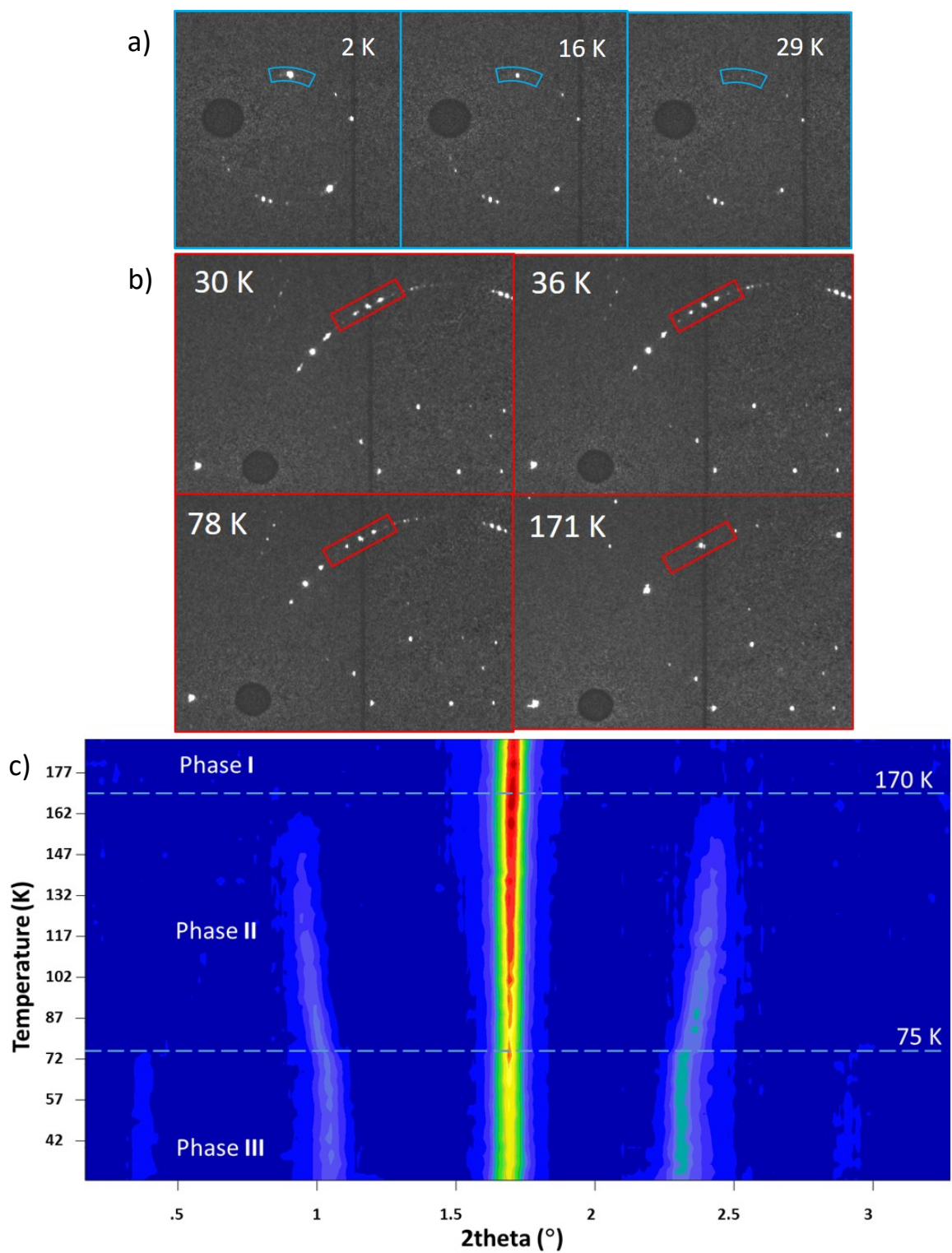

Figure S2. (a-b) Laue patterns of compound **1** collected at different temperatures in the CYCLOPS diffractometer at ILL. Below 170 K new satellite reflections appear indicating the phase transition from phase **I** to phase **II**. Below 75 K these satellite reflections become closer to the main reflection because a second phase transition to phase **III** with smaller **q** wave vector. Below 17 K new magnetic reflections

appear revealing the onset of magnetic order. Note that during the warming process, at 30 K, the sample was reoriented to investigate a region of reciprocal space where satellite reflections are more visible. (**c**) Mesh plot of the temperature evolution that corresponds with the rectangle marked in red in the patterns between 30 K and 190 K.

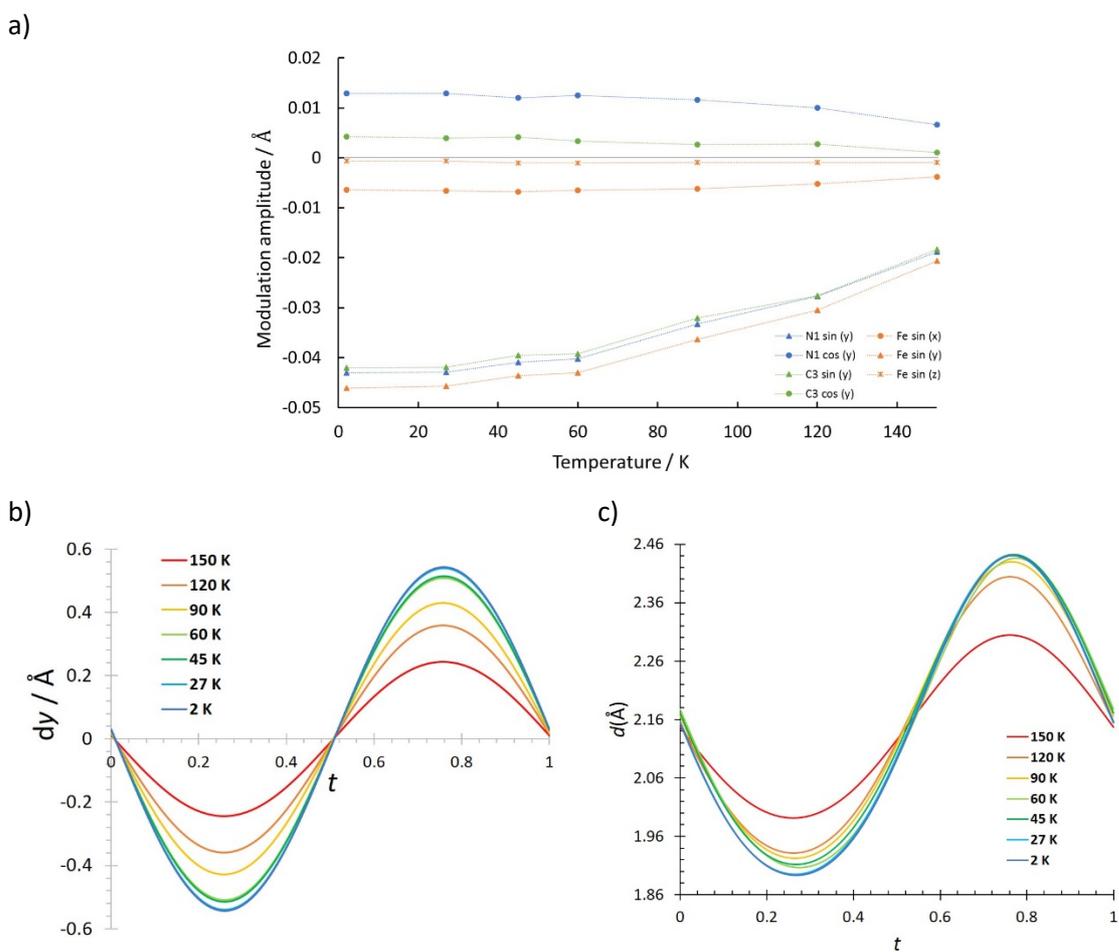

**Figure S3**. (**a**) The refined amplitude displacements in phases **II** and **III** for Fe(II) (orange), the C (green) and N (blue) atoms of the $(CH_3NH_3)^+$ counterion, representing the framework and the guest molecule, respectively. (**b**) Displacement along *b* axis of the iron atom and (**c**) H1N···O3 bond distances in the modulated phases of compound **1** where it can be seen the effect of temperature.

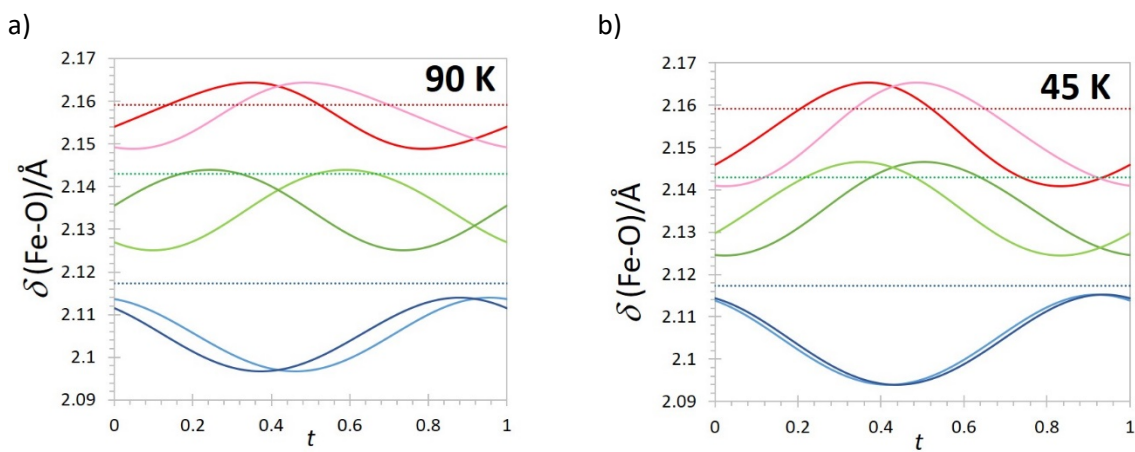

**Figure S4**. Modulation of the bond lengths between the iron and the oxygen atoms at 90 K (**a**) and 45 K (**b**) corresponding to phases **II** and **III**, respectively. The distances Fe1-O1, Fe1-O1a, Fe1-O2b, Fe1-O2c, Fe1-O3 and Fe1-O3a are represented in light blue, orange, grey, yellow, purple and green continuous lines, respectively. The distances Fe1-O1, Fe1-O2b and Fe1-O3 in the commensurate phase **I** are represented in blue, red and green dotted lines, respectively. *Symmetry code:* $a = -x, -y+1, -z+1$; $b = -x+1/2, -y+1, z-1/2$; $c = x-1/2, y, -z+3/2$.

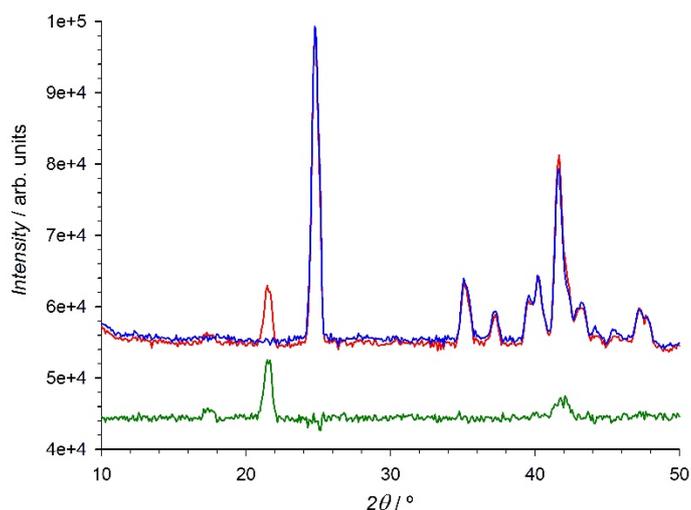

**Figure S5.** Neutron powder patterns of compound **1** collected at 2 K (red) and 26 K (blue) using the high flux D1B diffractometer. The difference diffraction pattern has been represented as a solid green line.

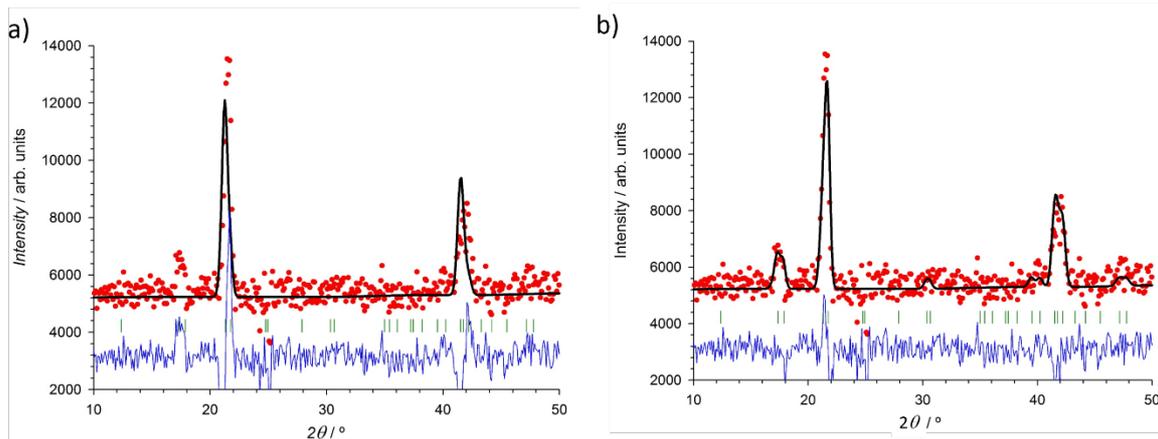

**Figure S6.** Fit of the difference pattern of compound **1** in the *Pn'ma'* and in the *Pnma* magnetic space group (**a** and **b**, respectively). Experimental data have been represented as red circles, calculated Rietveld patterns are shown as solid black lines, and difference between observed and calculated patterns have been plotted as solid blue lines. Vertical green marks represent the position of the Bragg reflections.

**Table S1.** List of the possible super space groups, obtained using ISODISTORT program,[1] compatible with the distortion modes obtained from considering the *Pnma* parent structure and **k** = (0, 0, 0) and $q_2$ = 0.1425(2)$c^*$ as modulation vectors for the magnetic and nuclear distortions, respectively.

| | |
|---|---|
| *Pnma*.1(00γ)000, origin = (0,0,0,0) | *Pnma*.1(00γ)0s0, origin = (0,0,0,0) |
| *Pmn2₁*.1(00γ)000, origin = (1/4,3/4,0,0) | *Pmn2₁*.1(00γ)s0s, origin = (1/4,3/4,0,0) |
| *Pn'm'a*(00γ)000, origin = (0,0,0,0) | *Pn'm'a*(00γ)0s0, origin = (0,0,0,0) |
| *Pm'n'2₁*(00γ)000, origin = (1/4,3/4,0,0) | *Pm'n'2₁*(00γ)s0s, origin = (1/4,3/4,0,0) |
| *Pnm'a'*(00γ)000, origin = (0,0,0,0) | *Pnm'a'*(00γ)0s0, origin = (0,0,0,0) |
| *Pm'n2₁'*(00γ)000, origin = (1/4,3/4,0,0) | *Pm'n2₁'*(00γ)s0s, origin = (1/4,3/4,0,0) |
| *Pn'ma'*(00γ)000, origin = (0,0,0,0) | *Pn'ma'*(00γ)0s0, origin = (0,0,0,0) |
| *Pmn'2₁'*(00γ)000, origin = (1/4,3/4,0,0) | *Pmn'2₁'*(00γ)s0s, origin = (1/4,3/4,0,0) |
| *Pnma*.1(00γ)0s0, origin = (0,0,0,1/4) | *Pnma*.1(00γ)000, origin = (0,0,0,1/4) |
| *Pmn2₁*.1(00γ)s0s, origin = (1/4,3/4,0,0) | *Pmn2₁*.1(00γ)000, origin = (1/4,3/4,0,0) |
| *Pn'm'a*(00γ)0s0, origin = (0,0,0,1/4) | *Pn'm'a*(00γ)000, origin = (0,0,0,1/4) |
| *Pm'n'2₁*(00γ)s0s, origin = (1/4,3/4,0,0) | *Pm'n'2₁*(00γ)000, origin = (1/4,3/4,0,0) |
| *Pnm'a'*(00γ)0s0, origin = (0,0,0,1/4) | *Pnm'a'*(00γ)000, origin = (0,0,0,1/4) |
| *Pm'n2₁'*(00γ)s0s, origin = (1/4,3/4,0,0) | *Pm'n2₁'*(00γ)000, origin = (1/4,3/4,0,0) |
| *Pn'ma'*(00γ)0s0, origin = (0,0,0,1/4) | *Pn'ma'*(00γ)000, origin = (0,0,0,1/4) |
| *Pmn'2₁'*(00γ)s0s, origin = (1/4,3/4,0,0) | *Pmn'2₁'*(00γ)000, origin = (1/4,3/4,0,0) |